\documentclass[12pt]{article}
\usepackage{graphicx}
\usepackage{amsfonts}
\usepackage{amssymb,amsmath}
\usepackage{latexsym}
\usepackage{color}
\usepackage{cite}
\input{colordvi.tex}

\newcommand{\del}{\partial}
\newcommand{\fnlmin}{-8.9}
\newcommand{\fnlmax}{14.3}

\begin{document}

\begin{titlepage}
\begin{flushright}
ICRR-Report:654-2013-3
\end{flushright}

\begin{center}

{\Large \bf 
Implications of Planck results for models with local type
non-Gaussianity}

\vskip .45in

{\large
Teruaki Suyama$^{1,2}$,
Tomo Takahashi$^3$, 
Masahide Yamaguchi$^4$,\\
and 
Shuichiro Yokoyama$^5$
}

\vskip .45in

{\em
$^1$Research Center for the Early Universe, Graduate School
  of Science, \\ The University of Tokyo, Tokyo 113-0033, Japan
 \vspace{0.2cm}\\
$^2$Leung Center for Cosmology and Particle Astrophysics (LeCosPA), 
National Taiwan University, Taipei 10617, Taiwan \vspace{0.2cm}\\
$^3$Department of Physics, Saga University, Saga 840-8502, Japan 
 \vspace{0.2cm} \\
$^4$Department of Physics, Tokyo Institute of Technology, Tokyo
152-8551, Japan
 \vspace{0.2cm} \\
$^5$Institute for Cosmic Ray Research,
The University of Tokyo\\
Kashiwa 277-8582, Japan

}

\end{center}

\vskip .4in

\begin{abstract}
We discuss implications of Planck results for models with local type
non-Gaussianity. In light of the recent results of the Planck satellite,
we constrain model parameters of several representative models and give
the prediction of trispectrum, in particular, $g_{\rm NL}$. We also
consider interesting possibilities that trispectrum appears as the first
signature of the non-Gaussianities of the curvature perturbations, that
is, $f_{\rm NL}$ is small while $g_{\rm NL}$ can be significantly large.
\end{abstract}

\end{titlepage}


\clearpage

\setcounter{page}{1}

\section{Introduction}

Very recently, the Planck mission has released data from the first 15.5
months of Planck operations for cosmic microwave background (CMB)
anisotropies \cite{PlanckI}. They determined cosmological parameters
with unprecedented accuracy such as the baryon, the dark matter, and the
dark energy densities, which strongly support the so-called concordance
model of cosmology \cite{PlanckXVI}. They also gave strong constraints
on primordial curvature perturbations. The spectral index $n_s = 0.9603
\pm 0.0073$ (68\%CL) \cite{PlanckXXII} significantly deviates from
unity, supporting the slow-roll inflation paradigm. Any deviations from
Gaussianities of primordial curvature perturbations are not found. In
particular, the local type of $f_{\rm NL}$ is now strongly constrained
as $\fnlmin < f_{\rm NL} < \fnlmax$ at two sigma level
\cite{PlanckXXIV}, which rules out a lot of light field models
predicting large local type non-Gaussianities.

In fact, one may wonder if light field models such as the curvaton
\cite{Enqvist:2001zp,Lyth:2001nq,Moroi:2001ct} and the modulated
reheating \cite{Dvali:2003em,Kofman:2003nx} scenarios might be excluded
because they are often claimed to generate large
non-Gaussianities. However, this is not the case. For example, as is
well known, in the curvaton scenario with a quadratic potential, the
local type of $f_{\rm NL}$ is given by \cite{Bartolo:2003jx,Lyth:2005du}
\begin{equation}
  f_{\rm NL} = \frac{5}{4r}-\frac53-\frac{5r}{6}.
\end{equation}
Here $r$ is roughly the fraction of the curvaton energy density at the
curvaton decay and is defined as
\begin{equation}
\left.  r = \frac{3\rho_{\sigma}}{3\rho_{\sigma}+4\rho_r} \right|_{\rm decay},
\end{equation}
where $\rho_{\sigma}$ is the curvaton energy density and $\rho_r$ is the
radiation energy density. %
Planck collaboration has reported the constraint on $r$ from a likelihood analysis
as $0.15 < r $ (95\% CL)  \cite{PlanckXXIV},
adopting a prior $0< r < 1$, 
which
rules out the curvaton model with small $r$.
However, to be
fair, a natural value of $r$ without fine-tuning is unity because the
curvaton easily dominates the energy density of the Universe since the
curvaton behaves like matter while the other components behave as
radiation. Such a value of $r(=1)$ yields $f_{\rm NL} = -5/4$, which is
still allowed by the recent Planck data. Thus, a simple and natural
model of the curvaton is still viable.

In the same way, the modulated reheating scenario predicts $f_{\rm NL}$
as \cite{Zaldarriaga:2003my,Ichikawa:2008ne}
\begin{equation}
  f_{\rm NL} = 5 \left( 1 - \frac{\Gamma \Gamma_{\sigma\sigma}}{\Gamma_{\sigma}^2} \right), 
\end{equation}
where $\Gamma$ is the inflaton decay rate depending on the modulus
$\sigma$, $\Gamma_{\sigma}=\partial \Gamma/\partial \sigma$,
$\Gamma_{\sigma\sigma}=\partial^2 \Gamma/\partial \sigma^2$, and the
inflaton is assumed to oscillate around its minimum with a quadratic
potential. 
Since the 
functional form of $\Gamma$ depends on the model, thus the 
parameter $\Gamma  \Gamma_{\sigma\sigma}/\Gamma_{\sigma}^2$
can be taken freely,
the constraint $\fnlmin < f_{\rm NL} < \fnlmax$  (95\% CL)  directly
leads to $-1.9 <
\Gamma \Gamma_{\sigma\sigma}/\Gamma_{\sigma}^2 < 2.8$  (95\% CL), which suggests that
the second derivative of $\Gamma$ with respect to $\sigma$ is
significantly suppressed. Thus, as long as $\Gamma$ linearly depends on
the modulus $\sigma$, $f_{\rm NL}$ is predicted to be 5, which is still
allowed by the recent Planck data.

Thus the light field models are still viable. After excluding the model
parameters inconsistent with Planck data, we end up with the light field
models that generically yield $f_{\rm NL}={\cal O}(1)$. Given that the
standard inflation model, in which the inflaton itself is responsible
for the curvature perturbations, predicts ${\cal O}(0.01)$\cite{Maldacena:2002vr,Creminelli:2004yq}, 
it is still important. Then, it is still important
to detect order of unity $f_{\rm NL}$ for discriminating light field
models from the standard inflation model though, in this case, we have
to take into account the intrinsic CMB bispectrum $f_{\rm NL} \lesssim 1$
coming from the second order effects of the evolution of the curvature
perturbations \cite{Pettinari:2013he,Huang:2012ub}.

In this paper, we first discuss how much light field models are
constrained according to the Planck data. As explained above, both of
the curvaton and the modulated scenarios are still allowed, and, in some
sense, the constrained value of $f_{\rm NL}$ is reasonable to avoid
fine-tuning. Then, we give constraints on model parameters of light
field models including the curvaton and the modulated scenarios and
predict the trispectrum, in particular $g_{\rm NL}$, for each model.
 
Next, we are going to pursue another interesting possibility. Though the
recent data of the Planck satellite claims that bispectrum of the
curvature perturbations is small, it does not necessarily imply that the
non-Gaussianities of the curvature perturbations are insignificant
because such non-Gaussianities might first appear on their
trispectrum. In the latter half of this paper, such a possibility will
be discussed in detail. Actually, for example, if we consider the
modulated reheating scenario and its modulus also behaves like a
curvaton, $f_{\rm NL}$ can be almost canceled and be small, but $g_{\rm
NL}$ still can be large. We are going to discuss such possibilities and
investigate which combination of model parameters can realize such
possibilities.

The organization of the paper is as follows. In the next section, we
constrain the model parameters of each light field model based on the
recent Planck results and give the predictions of the trispectrum
$g_{\rm NL}$, whose observations are essentially important for pinning
down the model. In Sec III. we discuss models, in which $f_{\rm NL}$ is
small, actually, within the constraints given by the Planck satellite,
but $g_{\rm NL}$ can be large as the first signal of the
non-Gaussianities of the curvature perturbations. Detailed explanations of
why $g_{\rm NL}$ can be large while keeping $f_{\rm NL}$ small for such
models are given. The final section is devoted to summary of this paper.

\section{Light field models}

During inflation, there can be light fields other than the
inflaton. They acquire quantum fluctuations during inflation, which can
be converted to the curvature perturbations. Though a lot of conversion
mechanisms have been proposed \cite{Suyama:2010uj}, the $\delta N$
formalism
\cite{Starobinsky:1986fxa,Salopek:1990jq,Sasaki:1995aw,Sasaki:1998ug,Lyth:2004gb}
%
based on the separate universe picture \cite{Kodama:1997qw,Nambu:1997wh}
enables us to make a systematic treatment to evaluate the final
curvature perturbations. According to the $\delta N$ formalism, the
superhorizon curvature perturbations at the final time $t=t_f$ can be
easily estimated by
\begin{equation}
 \zeta(t_f) = N_a \delta \varphi^a_\ast 
+ {1 \over 2}N_{ab}\delta \varphi^a_\ast \delta \varphi^b_\ast
 + {1 \over 6} N_{abc} 
 \delta  \varphi^a_\ast \delta \varphi^b_\ast \delta \varphi_\ast^c~,
\end{equation}
where $t_{\ast}$ is the time shortly after the horizon exit, a subscript
$a, b$, and $c$ represents a light field, $N_a = \del
N/\del\varphi^a_{\ast}$, and so on. Here, $\delta\varphi^a_{\ast}$
represents a field fluctuation evaluated at $t=t_{\ast}$ and is assumed
to be Gaussian. Then, the bispectrum and the trispectrum of the
curvature perturbations are characterized only by the three parameters,
$f_{\rm NL}^{\rm local}$, $g_{\rm NL}^{\rm local}$, and $\tau_{\rm
NL}^{\rm local}$ as follows,
\begin{eqnarray}
\langle \zeta_{\vec k_1} \zeta_{\vec k_2} \zeta_{\vec k_3} \rangle
&=&
{(2\pi)}^3 B_\zeta (k_1,k_2,k_3) \delta ({\vec k_1}+{\vec k_2}+{\vec
k_3}),
\nonumber \\
\langle
\zeta_{\vec k_1} \zeta_{\vec k_2} \zeta_{\vec k_3} \zeta_{\vec k_4}
\rangle
&=&
{(2\pi)}^3 T_\zeta (k_1,k_2,k_3,k_4) \delta ({\vec k_1}+{\vec k_2}+{\vec k_3}+{\vec k_4}),
\label{eq:bitri}
\end{eqnarray}
where
\begin{eqnarray}
B_\zeta (k_1,k_2,k_3)
&=&
\frac{6}{5} f_{\rm NL}^{\rm local}
\left(
P_\zeta (k_1) P_\zeta (k_2)
+ P_\zeta (k_2) P_\zeta (k_3)
+ P_\zeta (k_3) P_\zeta (k_1)
\right), \\
\label{eq:def_f_NL}
T_\zeta (k_1,k_2,k_3,k_4)
&=&
\tau_{\rm NL}^{\rm local} \left(
P_\zeta(k_{13}) P_\zeta (k_3) P_\zeta (k_4)+11~{\rm perms.}
\right) \nonumber \\
&&
+ \frac{54}{25} g_{\rm NL}^{\rm local} \left( P_\zeta (k_2) P_\zeta (k_3) P_\zeta (k_4)
+3~{\rm perms.} \right),
\label{eq:def_tau_g_NL}
\end{eqnarray}
with $k_{13} = |{\vec k_1} + {\vec k_3}|$. These three parameters are
easily evaluated at the tree level according to the $\delta N$
formalism,
\begin{eqnarray}
{6 \over 5}f_{\rm NL}^{\rm local}
&=& 
\frac{N_a N_b N^{ab}}{\left( N_c N^c \right)^2}, \nonumber \\
\tau_{\rm NL}^{\rm local}
&=& 
\frac{N_a N_{b} N^{ac} N_c^{~b}}{\left( N_d N^d \right)^3}, \nonumber \\
\frac{54}{25}g_{\rm NL}^{\rm local}
&=&  
\frac{N_{abc} N^a N^b N^c}{\left( N_d N^d \right)^3}.
\end{eqnarray} 
Below we omit the suffix ``local'' for simplicity. There is no general
relation between $f_{\rm NL}$ and $g_{\rm NL}$ because $g_{\rm NL}$
depends on the third derivative of $N$ in addition to the first one. On the other hand, there is a general inequality between
$f_{\rm NL}$ and $\tau_{\rm NL}$ \cite{Suyama:2007bg},
\begin{equation} 
  \tau_{\rm NL} \ge \left( \frac65 f_{\rm NL} \right)^2,
\end{equation}
When only one source (field) contributes to the curvature perturbations, the equality
must hold. On the other hand, when there are multiple sources, the
equality can be violated. In the following analysis, we concentrate on
single source case for simplicity. In this case, $\tau_{\rm NL}$ is
completely determined by $f_{\rm NL}$ as $\tau_{\rm NL} = 36 f_{\rm
NL}^2/25$. Therefore, $g_{\rm NL}$ is a key observable quantity to
discriminate light field models. Below, we constrain model parameters of
light field models and give the prediction for $g_{\rm NL}$ based on the
recent Planck satellite results. The extension of our discussions to
multiple source case is straightforward and, generally speaking, the
allowed region of the model parameters is widened.

\subsection{Curvaton}

In the curvaton model, the curvaton fluctuations are converted into the
curvature perturbations when the curvaton decays into relativistic
degrees of freedom, which occurs after inflation.  The important
quantities determining the resultant curvature perturbation are $r$ representing the curvaton fraction to the total energy density and
$\sigma_{\rm osc}$, the curvaton value when the curvaton starts
oscillations.  If a curvaton potential deviates from a quadratic form,
$\sigma_{\rm osc}$ generally depends on the curvaton value $\sigma_*$ at
the time of horizon crossing.  The non-linearity parameters from the
curvaton are given by \cite{Sasaki:2006kq}
\begin{eqnarray}
&&\frac{6}{5}f_{\rm NL}=\frac{3}{2r} \left( 1+\frac{\sigma_{\rm osc} \sigma''_{\rm osc}}{\sigma'^2_{\rm osc}} \right)-2-r, \\
&&\frac{54}{25}g_{\rm NL}=\frac{9}{4r^2} \left( \frac{\sigma_{\rm osc}^2 \sigma'''_{\rm osc}}{\sigma'^3_{\rm osc}}
+3 \frac{\sigma_{\rm osc} \sigma''_{\rm osc}}{\sigma'^2_{\rm osc}} \right)
-\frac{9}{r} \left( 1+\frac{\sigma_{\rm osc} \sigma''_{\rm osc}}{\sigma'^2_{\rm osc}} \right) \nonumber \\
&&\hspace{20mm}+\frac{1}{2} \left( 1-9\frac{\sigma_{\rm osc} \sigma''_{\rm osc}}{\sigma'^2_{\rm osc}} \right)+10r+3r^2,
\end{eqnarray}
where $\sigma_{\rm osc}' \equiv \partial \sigma_{\rm osc} /\partial
\sigma_*$ etc. Though the relation between $\sigma_{\rm osc}$ and
$\sigma_\ast$ is nontrivial in general, $\sigma_{\rm osc}$ has a
linear dependence on
%
 $\sigma_\ast$
 %
 for a quadratic potential of the
curvaton. In this case ($\sigma_{\rm osc}''=\sigma_{\rm osc}'''=0$), the
non-linear parameters reduce to
\begin{eqnarray}
&&\frac{6}{5}f_{\rm NL}=\frac{3}{2r} -2-r, \\
&&\frac{54}{25}g_{\rm NL}= -\frac{9}{r}+\frac{1}{2}+10r+3r^2,
\label{eq:gnlc}
\end{eqnarray} 
which lead to the following consistency relation,
\begin{equation}
 g_{\rm NL} = \frac{1}{54} \left[
               54 f_{\rm NL}^2 - 60 f_{\rm NL} - 125
               -\left( 9 f_{\rm NL}+5\right) 
                \sqrt{36 f_{\rm NL}^2+120f_{\rm NL}+250}  
\right].
\label{eq:gfc}
\end{equation} 
%
%
In this model, $r$ should be in the range of $0 < r < 1$.
As mentioned in the introduction, 
a likelihood analysis of $r$
with adopting a prior $0< r < 1$ gives the constraint $0.15 < r $ (95\% CL)  \cite{PlanckXXIV}.
Due to the fact that $r$ should be $r<1$, the non-linearity parameter $g_{\rm NL}$
is limited as $g_{\rm NL} < 2$. Furthermore, 
the Planck constraint can be translated into a lower bound for $g_{\rm NL}$
as  $-26.8 < g_{\rm NL} $ (95\% CL).

\subsection{Modulated reheating}

In the modulated reheating model, the decay rate of the inflaton
$\Gamma$ depends on some light field called modulus $\sigma$. After
inflation, the inflaton starts the oscillation around its minimum. When
the potential around the minimum is well approximated by a quadratic
type, the energy density of the inflaton oscillation decays in
proportional to $a^{-3}$ ($a$: the scale factor) and hence the
oscillation behaves like a non-relativistic matter. On the other hand,
the energy density of the Universe after reheating is dominated by
radiation, whose energy density decays in proportional to
$a^{-4}$. Thus, the fluctuations of the decay rate of the inflaton leads
to the energy density (curvature) perturbations of the Universe. 
Assuming that the decay rate is much smaller than the Hubble parameter evaluated at the
end of inflation \footnote{
This assumption does not necessarily hold in general.
For instance, see \cite{Watanabe:2013lwa}.}, 
the final curvature perturbation is easily evaluated by the $\delta N$ formalism as
\begin{eqnarray}
\zeta = 
-\frac16 \frac{\Gamma_{\sigma}}{\Gamma} \delta \sigma_\ast
+ \frac{1}{12} \left( 
-\frac{\Gamma_{\sigma \sigma}}{\Gamma} 
+ 
\frac{\Gamma_{\sigma}^2}{\Gamma^2} 
 \right) \delta \sigma_\ast^2
+ \frac{1}{36} \left( 
-\frac{\Gamma_{\sigma \sigma \sigma}}{\Gamma} 
+ 3 \frac{\Gamma_{\sigma} \Gamma_{\sigma \sigma}}{\Gamma^2}
-2 \frac{ \Gamma_{\sigma}^3}{\Gamma^3}
\right)\delta \sigma_\ast^3,
\end{eqnarray}
which yields the non-linear parameters,
\begin{eqnarray}
&&\frac{6}{5}f_{\rm NL} = 
6 
 -6 \frac{\Gamma \Gamma_{\sigma \sigma}}{\Gamma_\sigma^2}, \\
\notag  \\
&&\frac{54}{25}g_{\rm NL}
=
36 \left(
2 -3  \frac{ \Gamma \Gamma_{\sigma \sigma}}{\Gamma_\sigma^2}
 + \frac{\Gamma^2 \Gamma_{\sigma \sigma \sigma}}{\Gamma_\sigma^3}
 \right).
\end{eqnarray}
It should be noticed that, when $\Gamma$ linearly depends on $\sigma$
and the other higher derivatives vanish, $f_{\rm NL}$ and $g_{\rm NL}$
are predicted to be $f_{\rm NL} = 5$ and $g_{\rm NL} = 100/3$.

Let us consider the following $\sigma$ dependence on $\Gamma$ as a
more concrete example \cite{Ichikawa:2008ne},
\begin{equation}
\label{eq:Gamma_sigma}
\Gamma = \Gamma_0 \left( 
1 + \alpha \frac{\sigma}{M} + \beta \frac{\sigma^2}{M^2} 
\right),
\end{equation}
where $\alpha$ and $\beta$ are constants and M is some energy scale. The
non-linear parameters are rewritten as
\begin{equation}
\frac65 f_{\rm NL} \simeq 6 \left( 1 - \frac{2\beta}{\alpha^2} \right),
~~~~~
\frac{54}{25} g_{\rm NL} \simeq  36  \left( 2 - \frac{6\beta}{\alpha^2} \right).
\end{equation}
%
Since we do not need to assume any theoretical priors for the model parameters $\alpha$ and $\beta$,
the present constraint $\fnlmin < f_{\rm NL} < \fnlmax$ (95\% CL) directly
leads to $-0.9 < \beta/\alpha^2 < 1.4$ (95\% CL). 
In this example, the
third derivative of $\Gamma$, $\Gamma_{\sigma\sigma\sigma}$, is
negligible, which leads to the following consistency relation,
\begin{equation}
g_{\rm NL} = 10 f_{\rm NL} - \frac{50}{3}.
\end{equation}
Therefore, $g_{\rm NL}$ is predicted to be $-106 < g_{\rm NL} < 126$
(95\% CL).

\subsection{Inhomogeneous end of hybrid inflation}
\label{sec:hybrid}

In hybrid inflation, the inflationary phase are kept thanks to the positive
effective mass squared $m_{\chi}^2$ of the waterfall field $\chi$ and
ends at the critical value $\phi_{\rm cr}$ of the inflaton $\phi$ due to
the tachyonic instabilities. Then, if the effective mass squared of the
waterfall field depends not only on the inflaton but also on another
light field $\sigma$, the critical value $\phi_{\rm cr}$ also
fluctuates, which leads to the perturbation of the duration of the
inflation, that is, $\delta N$
\cite{Bernardeau:2002jf,Bernardeau:2004zz,Lyth:2005qk,Salem:2005nd,Alabidi:2006wa}.

Following Ref. \cite{Lyth:2005qk}, let us consider a potential of the form,
\begin{eqnarray}
V={\lambda \over 4}\left( \frac{v^2}{\lambda} - \chi^2\right)^2 + {1 \over 2}g^2\phi^2\chi^2 + {1 \over 2}m_\phi^2 \phi^2
+ {1 \over 2}f^2\sigma^2\chi^2 + {1 \over 2}m_\sigma^2 \sigma^2,
\end{eqnarray}
which yields the effective mass squared of the waterfall field $m_{\chi}^2$
and the critical value of the inflaton $\phi_{\rm cr}$ as follows,
\begin{equation}
  m_\chi^2 = -  v^2 + g^2 \phi^2 + f^2 \sigma^2,\qquad
  \phi_{\rm cr} = {\sqrt{v^2 - f^2 \sigma^2} \over g}.
\label{eq:masscri}
\end{equation}
Here $f, g$, and $\lambda$ are coupling constants and $v$ is some scale
related to the vacuum expectation value. The total duration (e-folding
number) of the inflation is easily estimated as
\begin{eqnarray}
N = - {1 \over M_{\rm Pl}^2}\int^{\phi_{\rm cr}}_{\phi_*} {V \over V_\phi} d\phi,
\end{eqnarray}
which fluctuates due to the perturbation of $\phi_{cr}$ and generates
the curvature perturbations,
\begin{eqnarray}
\zeta &=& {\partial N \over \partial \phi_{\rm cr}} {d \phi_{\rm cr} \over d\sigma}
\delta \sigma_*+{1 \over 2}\left[ 
{\partial^2 N \over \partial \phi_{\rm cr}^2} \left({d \phi_{\rm cr} \over d\sigma}\right)^2 +
{{\partial N \over \partial \phi_{\rm cr}} {d^2 \phi_{\rm cr} \over d\sigma^2}}\right] \delta \sigma_*^2~\nonumber\\
&&+{1 \over 6}\left[
{\partial^3 N \over \partial \phi_{\rm cr}^3} \left({d \phi_{\rm cr} \over d\sigma}\right)^3
+ 3 {\partial^2 N \over \partial \phi_{\rm cr}^2} \left({d \phi_{\rm cr} \over d\sigma}\right)\left({d^2 \phi_{\rm cr} \over d\sigma^2} \right)
+
{{\partial N \over \partial \phi_{\rm cr}} {d^3 \phi_{\rm cr} \over d\sigma^3}}\right] \delta \sigma_*^3.
\end{eqnarray}
Then, the non-linear parameters are given by
\begin{eqnarray}
{6 \over 5}f_{\rm NL} &\! \simeq \! &  - M_{\rm Pl} \sqrt{2\epsilon_{\rm cr}}
 {\phi''_{\rm cr} \over {\phi'_{\rm cr}}^2}, \\
{54 \over 25}g_{\rm NL} &\! \simeq \!&
-(2\epsilon_{\rm cr} - \eta_{\rm cr}) {18 \over 5}f_{\rm NL}
+ 2M_{\rm Pl}^2\epsilon_{\rm cr} {\phi'''_{\rm cr} \over {\phi'_{\rm cr}}^3}\label{eq:slowrollgnl},
\end{eqnarray}
where $\epsilon = M_{\rm pl}^2 (V_{\phi}/V)^2/2$ and $\eta = M_{\rm
pl}^2 V_{\phi\phi}/V$ are the standard slow-roll parameters and we have
omitted the contributions comparable to the slow-roll suppressed
parameters. The subscript ``cr'' represents the values evaluated at the
critical point $\phi_{\rm cr}$. From Eq. (\ref{eq:masscri}), we can
easily evaluate the non-linear parameters as
\begin{equation}
{6 \over 5}f_{\rm NL} = \eta_{\rm cr} { v^2 \over f^2 \sigma^2 },\qquad
{54 \over 25}g_{\rm NL} = 6 \eta_{\rm cr}^2 {v^2 \over f^2 \sigma^2 }.
\label{eq:inhomofg}
\end{equation}
%
Generically, the value of $\eta_{\rm cr}$
can be both positive and negative 
and hence there is no prior in this scenario, as in the modulated reheating case.
Then, the present constraint $\fnlmin < f_{\rm NL} < \fnlmax$ (95\% CL)
leads to $-10.7 < \eta_{\rm cr}v^2/(f^2 \sigma^2) < 17.2$ (95\% CL). We
also have the following consistency relation,
\begin{equation}
g_{\rm NL} = \eta_{\rm cr}\frac{10}{3} f_{\rm NL}.
\end{equation}
%
From Eq. (\ref{eq:inhomofg}) $g_{\rm NL}$ is theoretically bound as $g_{\rm NL} > 0$ and
since $|\eta_{\rm cr}| < 1$ $g_{\rm NL}$ is maximally predicted to be $g_{\rm NL} < 48$ (95\% CL).
%

\subsection{Inhomogeneous end of thermal inflation}
\label{sec:thermal}

In the inhomogeneous end of thermal inflation model
\cite{Kawasaki:2009hp}, the effective coupling between a flaton field
and the cosmic temperature, $g$, depends on some light field called
modulus $\sigma$.  During a mini-inflation phase, so-called thermal
inflation, the flaton field is trapped at the false vacuum due to the
thermal mass.  When the temperature decreases down to the critical
temperature depending on $g$, the flaton field starts to roll down to
its VEV and the mini-inflation ends.  Hence, the fluctuation of the
effective coupling $g$ leads to the fluctuations of cosmic $e$-folding
number corresponding to the primordial curvature perturbations.  The
final curvature perturbation is given by
\begin{equation}
\zeta = {1 \over 2} {g_\sigma \over g} \delta \sigma_*
+ {1 \over 4} \left( {g_{\sigma\sigma} \over g} - {g_\sigma^2 \over g^2}  \right) \delta \sigma_*^2
+ {1 \over 12} \left( {g_{\sigma\sigma\sigma} \over g}  
- 3 {g_{\sigma\sigma} g_\sigma \over g^2} + 2  {g_\sigma^3 \over g^3} \right) \delta \sigma_*^3, 
\end{equation}
where the subscript $\sigma$ denotes the derivative in terms of $\sigma$.
This yields the non-linearity parameters as
\begin{eqnarray}
&&\frac{6}{5}f_{\rm NL} = 
-2 
 +2 \frac{g g_{\sigma \sigma}}{g_\sigma^2}, \\
\notag  \\
&&\frac{54}{25}g_{\rm NL}
=
4\left(
2 - 3  \frac{ g g_{\sigma \sigma}}{ g_\sigma^2}
 + \frac{ g^2 g_{\sigma \sigma \sigma}}{ g_\sigma^3}
 \right).
\end{eqnarray}
Similarly to the modulated reheating case, when $g$ linearly depends on $\sigma$
and the other higher derivatives vanish, the non-linearity parameters are predicted to be
$f_{\rm NL} = - 5/ 3$ and $g_{\rm NL} = 25/9$.
When the effective coupling, $g$, has the following $\sigma$ dependence as
\begin{equation}
g = g_0 \left( 1 + \alpha {\sigma \over M} + \beta {\sigma^2 \over M^2} \right),
\label{eq:gform}
\end{equation}
the non-linearity parameters are given by
\begin{equation}
\frac65 f_{\rm NL} \simeq 2 \left( -1 + \frac{2\beta}{\alpha^2} \right),
~~~~~
\frac{54}{25} g_{\rm NL} \simeq  4  \left( 2 - \frac{6\beta}{\alpha^2} \right).
\end{equation}
%
Similarly to the modulated reheating case,
 the present constraint $\fnlmin < f_{\rm NL} < \fnlmax$
(95\% CL) leads to $-2.2 < \beta / \alpha^2 < 4.8$ (95\% CL) in this model.
We have also the consistency relation between the non-linearity parameters as
\begin{equation}
g_{\rm NL} = - {10 \over 3} f_{\rm NL} - {50 \over 27},
\end{equation}
where we have neglected $g_{\sigma\sigma\sigma}$ and this leads $-50 <
g_{\rm NL} < 28$ (95\% CL).

\subsection{Modulated trapping}

When the inflaton has a non-trivial coupling to other fields, the
resonant particle production can happen. Such particle production
significantly decreases the speed of the inflaton due to the
backreaction effects. Then, if such particle production process depends
on another light scalar field $\sigma$ through a coupling constant
and/or a resonant point, the cosmic expansion is perturbed due to the
perturbations of the light scalar field, which generates the curvature
perturbations.

Following Ref.~\cite{Langlois:2009jp}, let us consider the 
coupling between an inflaton $\phi$ and fermionic fields $\chi$ given by,
\begin{equation}
\mathcal{L}_{\rm int} = -\frac{1}{2} \mathcal{N} (m - \lambda \phi ) \bar{\chi} \chi,
\end{equation}
where $\mathcal{N}$ is the number of species of $\chi$ particles with
the same mass. Here, $m$ and $\lambda$ are a coupling constant and the
bare mass of $\chi$, both of which are assumed to depend on another
light field $\sigma$. When the inflaton reaches the particle production point
$\phi_{\rm pp} = m/\lambda$, the effective mass of $\chi$ vanishes so
that $\chi$ particles are resonantly produced. The produced number
density is estimated as
\begin{equation}
n_{\rm pp} = \frac{\lambda^{3/2}}{2 \pi^3} | \dot{\phi}_{\rm pp} |^{3/2},
\end{equation} 
where the subscript ``pp'' represents the quantities evaluated at the
particle production time. Then, the equation of motion for the inflaton
is modified as
\begin{equation}
\ddot{\phi} + 3 H \dot{\phi} + \frac{dV (\phi) }{d \phi} 
= 
\mathcal{N} \lambda n_{\rm pp}  \left( \frac{a}{a_{\rm pp}} \right)^{-3} \Theta (t  - t_{\rm pp}).
\end{equation}
In order to quantify the particle production effects, we define
\begin{equation}
\Delta \phi (t) \equiv \phi (t, \lambda \ne 0)  - \phi (t, \lambda=0), 
\end{equation} 
which is easily evaluated from the equation of motion,
\begin{equation}
\Delta \phi = \int_{t_\ast}^{\infty} \Delta \dot{\phi} dt 
= \frac{\mathcal{N} \lambda n_{\rm pp}}{9 H_{\rm pp}^2}.
\end{equation}
Here we regard $H$ and $dV(\phi)/d\phi$ almost constant because the
duration of particle production is assumed to be short. According to the
$\delta N$ formalism, the final curvature perturbations are estimated as
\begin{equation}
 \zeta = \Delta N^{(\lambda \ne 0 )} = - H_\ast \frac{\Delta \phi}{\dot{\phi}_{\rm pp}} 
= \frac{\lambda^{5/2} \mathcal{N} |\dot{\phi}_{\rm pp} |^{1/2}}{18 \pi^3
H_{\rm pp}},
\end{equation}
which can expanded with respect to $\delta\sigma_{\ast}$ as
\begin{equation}
\zeta 
= 
\left( \Delta N^{(\lambda \ne 0 )} \right)_{, \sigma} \delta \sigma_\ast 
+\frac{1}{2} \left( \Delta N^{(\lambda \ne 0 )} \right)_{, \sigma\sigma} (\delta \sigma_\ast)^2
+\frac{1}{6} \left( \Delta N^{(\lambda \ne 0 )} \right)_{, \sigma\sigma\sigma} (\delta \sigma_\ast)^3.
\end{equation}
Though the general formulae for the non-linear parameters are a bit
complicated, they are simply written in the case that both $m$ and
$\lambda$ are proportional to $\sigma$,
\begin{eqnarray}
\frac65 f_{\rm NL} &=& \frac{9}{5 e \beta}, \\ \notag \\
\frac{54}{25} g_{\rm NL} &=& \frac{27}{25 e^2 \beta^2}, 
\end{eqnarray}
which lead to the following consistency relation,
\begin{eqnarray}
g_{\rm NL} =  {2 \over 9} f_{\rm NL}^2.
\end{eqnarray}
Here, 
$e$ is so-called Euler's constant given by $e=2.7182 \dots $
 and $\beta$ is the ``efficiency factor'' given by
\begin{equation}
\beta \equiv 
\frac{ {\rm Max} (\Delta \dot{\phi}) }{ | \dot{\phi}_{\rm pp}|}
=
\frac{ \mathcal{N} \lambda^{5/2} |\dot{\phi}_{\rm pp} |^{1/2} } {6\pi^3 e H_{\rm pp}}.
\end{equation}
%
From this definition, $\beta$ should be in the range of $ 0 < \beta < 1$.
Then, a likelihood analysis of $\beta$ with adopting a prior $0 < \beta < 1$
gives the constraint $0.11 < \beta $ (95\% CL).
Similarly to the curvaton scenario,
the fact that $\beta < 1$ implies that $g_{\rm NL}$ should be $g_{\rm NL} > 0.06$
and the constraint  $0.11 < \beta $ (95\% CL)
predicts 
$ g_{\rm NL} < 5.6$ (95\% CL). 

\subsection{Velocity modulation}

In Ref. \cite{Nakayama:2011bc}, it was shown that if particles $\sigma$
in the early universe have velocity fluctuation on large scales, their
decay rate also acquires fluctuation through the fluctuation of the
Lorentz factor and hence the curvature perturbation is generated via the
same mechanism as the standard modulated reheating scenario.  One of the
scenario to realize such a situation is to assume that parent particle
$\Sigma$ which decays into the daughter particle $\sigma$ has
fluctuation of its mass $\delta m_\Sigma$ due to its dependence on the
light field having long wavelength fluctuations.  Notice that because of
the mass fluctuation $\delta m_\Sigma$, the resultant $\sigma$ particles
receive not only velocity modulation but also density perturbation
$\delta \rho_\sigma$.  As a result, the curvature perturbation generated
at the time when $\sigma$ decays consists of two components, namely, the
one generated by the standard curvaton mechanism and the other generated
by the velocity modulation.  Since both two components originate from
the same fluctuations $\delta m_\Sigma$, they are fully correlated to
each other.  Therefore, taking both the two effects mentioned above into
account, the final curvature perturbation can be expanded in terms of
$\delta m_\Sigma$ as
\begin{equation}
\zeta=A_1 \frac{\delta m_\Sigma}{m_\Sigma}
	+\frac{1}{2} A_2 {\left( \frac{\delta m_\Sigma}{m_\Sigma} \right)}^2+\frac{1}{6}A_3  {\left( \frac{\delta m_\Sigma}{m_\Sigma} \right)}^3,
\end{equation}
where each coefficient $A_i (i=1,2,3)$ depends on the four parameters in the model,
$w_0$ (equation of state parameter of $\sigma$ at the time of its generation),
$\Omega_\Sigma$ (density parameter of $\Sigma$ at the time of its decay),
$w_\sigma$ (equation of state parameter of $\sigma$ at the time of its decay) 
and $\Omega_\sigma$ (density parameter of $\sigma$ at the time of its decay).
The expressions for the expansion coefficients $A_i$ are lengthy (especially for $A_3$),
we refer the readers to Ref.~\cite{Nakayama:2011bc} for their explicit expressions. 
Assuming $w_0=\frac{1}{3}$ and $\Omega_\Sigma \ll 1$ just for simplicity,
the non-linearity parameters are given by
\begin{eqnarray}
f_{\rm NL}=&&\frac{20}{9 (3 w_\sigma (4 w_\sigma-1)+1)^2 \Omega_\sigma  ((3
   w_\sigma-1) \Omega_\sigma +4)}  \bigg[ w_\sigma (36 w_\sigma (2 w_\sigma-1)+5)+1) (\Omega_\sigma -3 w_\sigma \Omega_\sigma )^2 \nonumber \\
   &&+2 (w_\sigma (3 w_\sigma (24 w_\sigma (6 w_\sigma-7)+35)+10)-7) \Omega_\sigma +2 w_\sigma (9 w_\sigma (8 w_\sigma (6 w_\sigma+5)-21)+22)+22 \bigg], \label{vm-fnl} \\
g_{\rm NL}=&&\frac{800}{243 (3 w_\sigma (4 w_\sigma-1)+1)^3 \Omega_\sigma ^2 (-3
   w_\sigma \Omega_\sigma +\Omega_\sigma -4)^2}   \nonumber \\
 \times &&\bigg[ (3 w_\sigma-1) (w_\sigma (3 w_\sigma (27 w_\sigma (4 w_\sigma (96 w_\sigma (2
   w_\sigma-3)+127)-75)-43)-97)+41) \Omega_\sigma ^3  \nonumber \\ 
   &&+3 (w_\sigma (3 w_\sigma (9 w_\sigma (w_\sigma (12
   w_\sigma (8 w_\sigma (36 w_\sigma+19)-413)+2579)-416)+262)-500)+133) \Omega_\sigma ^2 \nonumber \\
   &&+2 (w_\sigma (36
   w_\sigma (3 w_\sigma-1) (4 w_\sigma-1)+1)+1) (\Omega_\sigma -3 w_\sigma \Omega_\sigma )^4 \nonumber \\
   &&+2 (w_\sigma (3
   w_\sigma (3 w_\sigma (3 w_\sigma (12 w_\sigma (12 (37-12 w_\sigma) w_\sigma+133)-5177)+4570)-728)+674)-517) \Omega_\sigma \nonumber \\
   && +4 (w_\sigma+1) (9 w_\sigma (w_\sigma (12 w_\sigma (36
   w_\sigma (4 w_\sigma+9)-181)+265)-2)+209)\bigg]. \label{vm-gnl}
\end{eqnarray}
We can easily find that $f_{\rm NL}$ is at least ${\cal O}(1)$ for any choice of
$(w_\sigma,~\Omega_\sigma)$.  For fixed value of $\Omega_\sigma$,
$f_{\rm NL}$ takes a maximum $\sim 15/\Omega_\sigma$ for $w \simeq 0.1$.
Numerically, we find that $0.5 \lesssim g_{\rm NL}/\tau_{\rm NL} \leq 1$
for any $(w_\sigma,~\Omega_\sigma)$.  Thus, a consistency relation
between $f_{\rm NL}$ and $g_{\rm NL}$ can be written as
\begin{equation}
g_{\rm NL}=C_{\rm vm} \frac{36}{25}f_{\rm NL}^2,~~~~~~~0.5 \lesssim C_{\rm vm} \leq 1.
\end{equation}
Using the Planck result $\fnlmin < f_{\rm NL} < \fnlmax$ (95\% CL), the
most conservative limit on $g_{\rm NL}$ in this model is given by
$0<g_{\rm NL}<294$ (95\% CL).

\section{Trispectrum as first signature of non-Gaussianity}

In this section, we discuss interesting possibility in which trispectrum
appears as a first signature of non-Gaussianities of the curvature
perturbations. That is, we discuss the cases that $f_{\rm NL}$ is small,
as shown by the Planck results, but $g_{\rm NL}$ can be significantly
large. Note again that $\tau_{\rm NL}$ is also small for single source
because we have the relation $\tau_{\rm NL}=36 f_{\rm NL}^2/25$.

Generally speaking, in order to realize large $g_{\rm NL}$ while keeping
$f_{\rm NL}$ small (that is, $N_{\sigma\sigma}/N_{\sigma}^2 = {\cal
O}(1),\,\, N_{\sigma\sigma\sigma}/N_{\sigma}^3 \gg 1$), we can take two
options. The first option is to consider large third derivative of $N$,
that is, $N_{\sigma\sigma\sigma}$. Such examples include curvaton with a
self coupling, modulated reheating model with the non-trivial(cubic)
dependence of a decay rate $\Gamma$ on the modulus $\sigma$, and
inhomogeneous end of thermal inflation with the cubic dependence of a
coupling $g$ on the modulus $\sigma$. In these examples, as long as such
cubic dependences are extraordinarily large, $g_{\rm NL}$ can be
significantly large, which is quite manifest from the formulae given in
the previous section.

The second option is to make the first derivative $N_{\sigma}$
accidentally small,\footnote{It should be noted that we also need to 
take into account the normalization of the power spectrum of the curvature
perturbations.} which leads to large $g_{\rm NL}$ unless
$N_{\sigma\sigma\sigma}$ is also suppressed. It should be noticed that
the second derivative $N_{\sigma\sigma}$ needs to be mildly suppressed
to keep $f_{\rm NL}$ of the order of unity. Such accidentally small
$N_{\sigma}$ can be obtained by tuning model parameters. For example, in
the modulated decay of the curvaton model
\cite{Langlois:2013dh,Assadullahi:2013ey,Enomoto:2013qf}, (essentially)
two model parameters ($r$ and $\Gamma
\Gamma_{\sigma\sigma}/\Gamma_{\sigma}^2$) appear. By taking their
adequate combination, $N_{\sigma}$ ($N_{\sigma\sigma}$) is significantly
(mildly) suppressed, which leads to large $g_{\rm NL}$ with small
$f_{\rm NL}$ without resorting to $N_{\sigma\sigma\sigma}$.

Such a situation also happens in the case that the same light field
$\sigma$ contributes to the curvature perturbations multiple times in
different ways. For example, in light field models discussed in the
previous section, the fluctuations of a light field other than the
inflaton are converted to the curvature perturbations. However, except
the curvaton (and the velocity modulation), the evolution of such a
light field is followed only until the conversion and the subsequent
evolution is simply assumed to be negligible. However, generally
speaking, such a modulus has only weak (gravitationally suppressed)
interactions so that it is long-lived and can easily contribute to the
energy density of the Universe at late times. That is, it is probable
that a modulus in a light field model also behaves like a curvaton at
late epoch, which becomes another origin of the curvature perturbations.
In this setting, the curvature perturbations $\zeta$ consists of two
parts. The first part comes from each light field model contribution and
the second part arises from the curvaton contribution,
\begin{equation}
  \zeta = \zeta_{\rm light}+\zeta_{\rm cur}.
  \label{eq:lightcurv}
\end{equation}
It should be noticed that both contributions $\zeta_{\rm light}$ and
$\zeta_{\rm cur}$ arise only from the same perturbations $\delta
\sigma_{\ast}$. Then, when the linear and quadratic terms cancel
adequately but the cubic term does not, $f_{\rm NL}$ becomes small while
significantly large $g_{\rm NL}$ can appear without resort to the cubic
dependence of the couplings. This is exactly the second option we can
take.

In this section, we discuss this kind of possibilities in detail. More
concretely, we investigate, for each model, which conditions on model
parameters are necessary to realize such a possibility.

\subsection{Modulated Decay of the Curvaton}

Now in this section, we consider a model where the decay rate of the
curvaton is modulated due to fluctuations of some other light scalar
field $\sigma$
\cite{Langlois:2013dh,Assadullahi:2013ey,Enomoto:2013qf}. In this model,
the curvature perturbation, up to the third order, is given by
\begin{eqnarray}
\label{zeta3_mod}
\zeta&=&-\frac{r}{6}\frac{\Gamma_\sigma}{\Gamma}\delta \sigma_\ast
-\frac{r}{72}  
\left[6 \frac{\Gamma_{\sigma\sigma}}{\Gamma} +\left(r^2+2 r-9\right) \frac{\Gamma_\sigma^2}{\Gamma^2} \right] \delta \sigma_\ast^2 
\notag \\
&&
-\frac{r}{1296} \left[36 \frac{\Gamma_{\sigma\sigma\sigma}}{\Gamma} 
 +18 \left(r^2+2 r-9\right)
   \frac{ \Gamma_\sigma \Gamma_{\sigma\sigma}}{\Gamma^2} 
      +\left(3 r^4+10 r^3-22 r^2-54
   r+135\right) \frac{\Gamma_\sigma^3}{\Gamma^3}\right]\delta \sigma_\ast^3\,, \notag \\
\end{eqnarray}
where $r$ is the fraction of the energy density of the curvaton at the
time of its decay and $\Gamma_\sigma = d \Gamma / d \sigma$ and so on.
In this model, the non-linearity parameters are written as
\begin{eqnarray}
f_{\rm NL} & = &\frac{5}{2r} \left(3-2 \frac{ \Gamma\Gamma_{\sigma\sigma} }{\Gamma_\sigma^{2}}\right) -2-r,
\\ \notag \\
g_{\rm NL}
 & = &  
 \frac{25}{54 r^2} \left[
 36 \frac{ \Gamma^2 \Gamma_{\sigma\sigma\sigma}}{\Gamma_\sigma^3} 
 + 18 \left(r^2+2 r-9\right) \frac{ \Gamma  \Gamma_{\sigma\sigma}}{\Gamma_\sigma^2} +3 r^4+10 r^3-22 r^2-54 r+135
  \right]. \notag \\ 
\end{eqnarray}
When $r \ll 1$, $f_{\rm NL}$ and $g_{\rm NL}$ are related as
\begin{equation}
g_{\rm NL} =\frac23 
\left( 15 + 4  \frac{\Gamma^2 \Gamma_{\sigma\sigma\sigma}}{\Gamma_\sigma^3}  -18 \frac{ \Gamma  \Gamma_{\sigma\sigma}}{\Gamma_\sigma^2} \right)
\left( - 2 \frac{ \Gamma  \Gamma_{\sigma\sigma}}{\Gamma_\sigma^2}  + 3 \right)^{-2}  f_{\rm NL}^2. 
\end{equation}
When the combination $ \Gamma \Gamma_{\sigma\sigma} / \Gamma_\sigma^2 $
is almost tuned to cancel the denominator in the coefficient of $f_{\rm
NL}^2$ in the right hand side, $g_{\rm NL}$ can be large even if $f_{\rm
NL}$ is $\mathcal{O}(1)$ since the numerator is not necessarily canceled
by such a choice of the functional form of $\Gamma$.

\subsection{Modulated curvaton}

Let us consider the case where the light field model contribution in Eq. (\ref{eq:lightcurv})
comes from the modulated reheating mechanism.
Assuming the quadratic potential of the light field and neglecting $\Gamma_{\sigma\sigma\sigma}$,
in such case the total curvature perturbation is given by
\begin{eqnarray}
\zeta 
&=&
\left( 
\frac{2r}{ 3\sigma_\ast}
-\frac{\Gamma_{\sigma}}{ 6 \Gamma}
\right ) \delta \sigma_\ast
+
\left[ 
\frac{1}{9 \sigma_\ast^2} \left( 3r - 4 r^2 -2  r^3
\right)
- \frac{1}{12} \left(
\frac{\Gamma_{\sigma \sigma}}{\Gamma} 
-
 \frac{\Gamma_{\sigma}^2}{\Gamma^2} 
 \right)
\right] \delta \sigma_\ast^2 \notag \\
&& 
+
\left[ 
\frac{4}{81\sigma_\ast^3} \left(
-9r^2
+\frac{r^3}{2} 
+10r^4 + 3r^5
\right)
-\frac{1}{36} \left( 
\frac{2 \Gamma_{\sigma}^3}{\Gamma^3}
- \frac{3 \Gamma_{\sigma} \Gamma_{\sigma \sigma}}{\Gamma^2}\right)
\right]
\delta \sigma_*^3. \notag \\
\end{eqnarray}
The non-linearity
parameters are then given by
\begin{eqnarray}
f_{\rm NL}
&=&
\frac{5}{3} {\left( 
-\frac{4r}{\sigma_\ast} +\frac{\Gamma_{\sigma}}{\Gamma} 
\right)}^{-2} 
\left[ 3 \left( \frac{\Gamma_{\sigma}^2}{\Gamma^2} 
-\frac{\Gamma_{\sigma \sigma}}{\Gamma} \right)
+\frac{4r}{\sigma_\ast^2} (3 - 4r - 2r^2) 
\right], \\ \notag \\
g_{\rm NL}
&=& 
- \frac{50}{3} {\left(  
-\frac{4r}{\sigma_\ast} 
+\frac{\Gamma_{\sigma}}{\Gamma} \right)}^{-3} 
\left[ 
-\frac{2 \Gamma_{\sigma}^3}{\Gamma^3}
+ \frac{3 \Gamma_{\sigma} \Gamma_{\sigma \sigma}}{\Gamma^2} +\frac{8r^2}{9 \sigma_\ast^3} \left( -18+r+20r^2+6r^3\right)
\right]. \notag \\
\end{eqnarray}
From the above expressions, we find that for fine-tuned parameters
we simultaneously have small $f_{\rm NL}$ and large $g_{\rm NL}$.
For example, in case where $\Gamma$ is given by Eq. (\ref{eq:Gamma_sigma})
and the parameters are set to be $\alpha=1$, $r=\Lambda^{-1}$ and $M/\sigma_* = \Lambda / 4$
with a big parameter $\Lambda \gg 1$,
we can have a situation where 
\begin{equation}
f_{\rm NL} \lesssim {\cal O}(1), \qquad
g_{\rm NL} = {\cal O}(\Lambda),
\end{equation}
by choosing $\beta$ appropriately.
Hence, we can simultaneously realize large $g_{\rm NL}$ and $f_{\rm NL} \lesssim {\cal O}(1)$.

\subsection{Inhomogeneous end of hybrid inflation and curvaton}

In a similar way to the above discussion about the modulated curvaton mechanism,
we consider the case where the fluctuations of the curvaton also induce inhomogeneous
end of hybrid inflation discussed in \ref{sec:hybrid}.
In this case, the total curvature perturbations are given by
\begin{eqnarray}
\zeta 
&=&
\left( 
\frac{2r}{ 3\sigma_\ast}+
\frac{1}{\eta_{\rm cr}} \frac{f^2\sigma_\ast}{g^2 \phi_{\rm cr}^2}
\right ) \delta \sigma_\ast \notag \\
&&
+
\left[ 
\frac{1}{9 \sigma_\ast^2} \left( 3r - 4 r^2 -2  r^3
\right)
-
\frac{1}{2M_{\rm Pl}^2} \frac{f^4 \sigma_\ast^2}{g^4 \phi_{\rm cr}^2}
+\frac{1}{2\eta_{\rm cr}}\frac{f^2 \left( v^2 + f^2 \sigma_\ast^2\right)}{g^4 \phi_{\rm cr}^4}
\right] \delta \sigma_\ast^2 \notag \\
&& 
+
\left[ 
\frac{4}{81\sigma_\ast^3} \left(
-9r^2
+\frac{r^3}{2} 
+10r^4 + 3r^5
\right)
-\frac{1}{6M_{\rm Pl}^2} \frac{f^4 \sigma_\ast (f^2\sigma_\ast^2 + 3 v^2)}{g^6 \phi_{\rm cr}^4}
+ \frac{1}{6\eta_{\rm cr}}
 \frac{f^4 \sigma_\ast (2f^2\sigma_\ast^2 + 6 v^2)}{g^6 \phi_{\rm cr}^6}
\right]
\delta \sigma_*^3. \notag \\
\end{eqnarray}
Neglecting the Planck suppressed terms, the non-linearity parameters are
given by
\begin{eqnarray}
f_{\rm NL}
&\simeq&
\frac{5}{3} {\left( 
\frac{2r}{3}+
\frac{1}{\eta_{\rm cr}} \frac{f^2\sigma_\ast^2}{g^2 \phi_{\rm cr}^2}
\right)}^{-2} 
\left[ \
\frac{1}{9} \left( 3r - 4 r^2 -2  r^3
\right)
+\frac{1}{2\eta_{\rm cr}}\frac{f^2 \sigma_\ast^2 \left( v^2 + f^2 \sigma_\ast^2\right)}{g^4 \phi_{\rm cr}^4}
\right], \\ \notag \\
g_{\rm NL}
&\simeq& 
 \frac{50}{9} {\left(  
\frac{2r}{3}+
\frac{1}{\eta_{\rm cr}} \frac{f^2\sigma_\ast^2}{g^2 \phi_{\rm cr}^2}
 \right)}^{-3} \notag \\
 && \times
\left[ 
\frac{4}{81} \left(
-9r^2
+\frac{r^3}{2} 
+10r^4 + 3r^5
\right)
+ \frac{1}{6\eta_{\rm cr}}
 \frac{f^4 \sigma_\ast^4 (2f^2\sigma_\ast^2 + 6 v^2)}{g^6 \phi_{\rm cr}^6}
\right]. \notag \\
\end{eqnarray}
In denominators in the above expressions, both terms are positive
definite and hence in order to realize large $g_{\rm NL}$ the both terms
must be much smaller than unity at least.  For $r < 1$, in the numerator of
the expression of $f_{\rm NL}$ it is not possible to realize a cancellation to
obtain the small $f_{\rm NL}$.  Hence, the small
$f_{\rm NL}$ does not yield $g_{\rm NL}$ large enough to be detected in this scenario.

\subsection{Inhomogeneous end of thermal inflation and curvaton}

We can also consider the case where the light field contribution comes from 
the inhomogeneous end of thermal inflation considered in \ref{sec:thermal}.
In such case, we have
\begin{eqnarray}
\zeta 
&=&
\left( 
\frac{2r}{ 3\sigma_\ast} + {1 \over 2}{g_\sigma \over g}
\right ) \delta \sigma_\ast
+
\left[ 
\frac{1}{9 \sigma_\ast^2} \left( 3r - 4 r^2 -2  r^3
\right)
+ {1 \over 4} \left( {g_{\sigma\sigma} \over g} - {g_\sigma^2 \over g^2}  \right)
\right] \delta \sigma_\ast^2 \notag \\
&& 
+
\left[ 
\frac{4}{81\sigma_\ast^3} \left(
-9r^2
+\frac{r^3}{2} 
+10r^4 + 3r^5
\right)
+
{1 \over 12} \left( 
- 3 {g_{\sigma\sigma} g_\sigma \over g^2} + 2  {g_\sigma^3 \over g^3} \right)
\right]
\delta \sigma_*^3. \notag \\
\end{eqnarray}
The non-linearity
parameters are then given by
\begin{eqnarray}
f_{\rm NL}
&=&
\frac{5}{3} {\left( 
\frac{4r}{3\sigma_\ast} +\frac{g_{\sigma}}{g} 
\right)}^{-2} 
\left[ \frac{4r}{9 \sigma_\ast^2} \left( 3 - 4 r -2  r^2
\right)
+  \left( {g_{\sigma\sigma} \over g} - {g_\sigma^2 \over g^2}  \right)
\right], \\ \notag \\
g_{\rm NL}
&=& 
 \frac{50}{27} {\left(  
\frac{4r}{3\sigma_\ast} 
+\frac{g_{\sigma}}{g} \right)}^{-3} 
\left[ 
\left( 
- 3 {g_{\sigma\sigma} g_\sigma \over g^2} + 2  {g_\sigma^3 \over g^3} \right) +\frac{8r^2}{27 \sigma_\ast^3} \left( -18+r+20r^2+6r^3\right)
\right]. \notag \\
\end{eqnarray}
Similar to the case of the modulated curvaton case, for example, in case where
$g$ is given by Eq. (\ref{eq:gform}) and the parameters are set to be
$\alpha=-1$, $r=\Lambda^{-1}$ and $M/\sigma_* = \Lambda / 4$
with a large parameter $\Lambda$, we have
\begin{equation}
f_{\rm NL} \lesssim {\cal O}(1), \qquad
g_{\rm NL} = {\cal O}(\Lambda),
\end{equation}
by choosing $\beta$ appropriately. Hence, we can simultaneously realize
large $g_{\rm NL}$ and $f_{\rm NL} \lesssim {\cal O}(1)$.

\subsection{Modulated trapping and curvaton}

In case where a light field inducing modulated trapping mechanism behaves like curvaton in
later dynamics,
the curvature perturbation is given by
\begin{eqnarray}
\zeta 
&=&
\left( 
\frac{2r}{ 3\sigma_\ast} + \frac{5 e \beta}{6 \sigma_\ast}
\right ) \delta \sigma_\ast
+
\left[ 
\frac{1}{9 \sigma_\ast^2} \left( 3r - 4 r^2 -2  r^3
\right)
+\frac{5e\beta}{8\sigma_\ast^2}
\right] \delta \sigma_\ast^2 \notag \\
&& \qquad\qquad
+
\left[ 
\frac{4}{81\sigma_\ast^3} \left(
-9r^2
+\frac{r^3}{2} 
+10r^4 + 3r^5
\right)
+
\frac{5 e\beta}{48 \sigma_\ast^3}
\right]
\delta \sigma_*^3. \notag \\
\end{eqnarray}
The non-linearity parameter are given by
\begin{eqnarray}
f_{\rm NL}
&=&
\frac{5}{3} {\left(4r + 5e\beta
\right)}^{-2} 
\left[ 4r \left( 3 - 4 r -2  r^2
\right)
+  \frac{45}{2} e\beta
\right], \\ \notag \\
g_{\rm NL}
&=& 
 \frac{25}{9} {\left(  
4r + 5 e\beta\right)}^{-3} 
\left[ \frac{16 r}{3}\left( -18 + r+20r^2 +6 r^3\right)
+ \frac{45}{2} e\beta
\right]. \notag \\
\end{eqnarray}

In this scenario, due to the constraints on the parameters $\beta$ and
$r$ as $0 < \beta < 1$ and $0<r<1$, it is hard to realize the large
$g_{\rm NL}$ without violating the Planck constraint on $f_{\rm NL}$.

\section{Summary}

Following the Planck 2013 results \cite{PlanckXXIV}, we discussed models
of generating local-type non-Gaussianity where a light field other than
the inflaton plays a main role of generating the curvature
perturbations.  First, we showed the constraint on model parameters for
each light field model, by introducing the constraint on local-type
$f_{\rm NL}$ obtained in Planck 2013 results: XXIV.  By using the
consistency relation between the non-linearity parameters $f_{\rm NL}$
and $g_{\rm NL}$ for light field models as given in our previous paper
\cite{Suyama:2010uj}, we also obtained the constraint on $g_{\rm NL}$
and found that for simple light field models $g_{\rm NL}$ does not
become large enough to be detected even in forthcoming Planck data
and other future experiments,
within the Planck constraint on $f_{\rm NL}$
\cite{Desjacques:2009jb,Smidt:2010ra,Regan:2010cn,Sekiguchi:2013hza,Giannantonio:2013uqa,Hikage:2012bs}.

We also discussed the possibility of generating large $g_{\rm NL}$ in
light field models within the Planck constraint on $f_{\rm NL}$.  We
classified the possible models into two categories. One is to consider
large third derivative of $N$, that is, $N_{\sigma\sigma\sigma}$ in
$\delta N$ formalism.  Such examples include curvaton scenario with a
self coupling and also modulated reheating (or inhomogeneous end of
thermal inflation) scenario with non-trivial cubic dependence of a decay
rate $\Gamma$ (or a coupling $g$) on the modulus $\sigma$ which does not
appear in the non-linearity parameter $f_{\rm NL}$.  In such case, the
consistency relation between $f_{\rm NL}$ and $g_{\rm NL}$ is no longer
realized and hence we can realize the large $g_{\rm NL}$ within the
Planck constraint.  Another is to make the first derivative $N_\sigma$
accidentally small, which leads to large $g_{\rm NL}$ unless
$N_{\sigma\sigma\sigma}$ is also suppressed.  Such accidentally small
$N_\sigma$ can be obtained in the case where the same light field
$\sigma$ contributes to the curvature perturbations multiple times in
different ways.  As examples, we considered the cases where a modulus in
a light field model also behaves like a curvaton at late epoch.  We
found that when we consider the modulated reheating or the inhomogeneous
end of thermal inflation scenario as a light field model by taking
fine-tuned parameters we can realize the large $g_{\rm NL}$ due to the
accidental cancellation.  However, for the cases where the inhomogeneous
end of hybrid inflation or the modulated trapping scenario is considered
as a light field model we found it difficult to realize accidentally
small $N_\sigma$ and $N_{\sigma\sigma}$ and hence obtaining large
$g_{\rm NL}$ within the Planck constraint is also difficult.  Although
information about the $g_{\rm NL}$ would be a useful tool to
discriminate the light field models in future observations, it seems to
be difficult to realize the measurable $g_{\rm NL}$ without fine-tuning.

\section*{Acknowledgments}

We would like to thank Nicola Bartolo for clarifying the constraint on
the curvaton parameter in the Planck paper
and also Jun'ichi Yokoyama for useful comments. TS thanks the Leung Center
for Cosmology and Particle Astrophysics (LeCosPA), National Taiwan
University for the kind hospitality during his visit when this paper is
completed. This work was supported in part by the Grant-in-Aid for
Scientific Research No.~1008477 (TS), No.~23740195 (TT), and
No.~21740187 (MY), the Grant-in-Aid for Scientific Research on
Innovative Areas No.~24111706 (MY), and Grant-in-Aid for JSPS Fellows
No. 24-2775(SY).

\end{document}